\documentclass[twocolumn,superscriptaddress,showpacs,prl,amsmath,amstex,amssymb,citeautoscript,longbibliography]{revtex4-1}
\pdfoutput=1
\usepackage{natbib}
\usepackage[english]{babel}
\usepackage{letltxmacro}
\usepackage{latexsym}
\usepackage{graphicx}
\usepackage{amsmath}
\usepackage{amsfonts}
\usepackage{amssymb}
\usepackage{bm}
\usepackage{color}
\usepackage[percent]{overpic}
\usepackage{soul} 
\usepackage{amssymb}
\usepackage{wasysym}
\usepackage{dsfont}
\usepackage{float}
\usepackage{graphicx}
\usepackage[colorinlistoftodos]{todonotes}
\usepackage{verbatim}
\usepackage[normalem]{ulem}
\usepackage{url}
\usepackage{breakurl}
\usepackage{hyperref}
\hypersetup{
	bookmarks=false,         
	unicode=false,          
	pdftoolbar=false,        
	pdfmenubar=true,        
	pdffitwindow=false,     
	pdfstartview={FitH},    
	pdftitle={},    
	pdfauthor={Authors},     
	pdfsubject={},   
	pdfcreator={},   
	pdfproducer={}, 
	pdfkeywords={many-body localization} {matrix elements} {disordered systems}, 
	pdfnewwindow=true,      
	colorlinks=true,       
	linkcolor=black,          
	citecolor=blue,        
	filecolor=magenta,      
	urlcolor=blue           
}

\newcommand{\Tr}{\mathrm{Tr}}

\newcommand{\beq}{\begin{eqnarray}}
\newcommand{\eeq}{\end{eqnarray}}

\usepackage{mathtools}

\DeclarePairedDelimiter{\floor}{\lfloor}{\rfloor}

\begin{document}

\title{Variational Thermal Quantum Simulation via Thermofield Double States}
\author{Jingxiang Wu}
\affiliation{Perimeter Institute for Theoretical Physics, 31 Caroline St.\,N., Waterloo, ON N2L 2Y5, Canada}
\affiliation{Department of Physics $\&$ Astronomy, University of Waterloo,
	 Waterloo, ON N2L 3G1, Canada}
\author{Timothy H. Hsieh}
\affiliation{Perimeter Institute for Theoretical Physics, 31 Caroline St.\,N., Waterloo, ON N2L 2Y5, Canada}
\date{\today}
\begin{abstract}

We present a variational approach for quantum simulators to realize finite temperature Gibbs states by preparing thermofield double (TFD) states.  Our protocol is motivated by the quantum approximate optimization algorithm (QAOA) and involves alternating time evolution between the Hamiltonian of interest and interactions which entangle the system and its auxiliary counterpart.  As a simple example, we demonstrate that thermal states of the 1d classical Ising model at any temperature can be prepared with perfect fidelity using $L/2$ iterations, where $L$ is system size.  We also show that a free fermion TFD can be prepared with nearly optimal efficiency.  Given the simplicity and efficiency of the protocol, our approach enables near-term quantum platforms to access finite temperature phenomena via preparation of thermofield double states.    

\end{abstract}
\maketitle

Rapid advances in the control and measurement of a variety of quantum platforms, such as trapped ions \cite{Blatt12, Zhang2017_DPT,Islam583} , superconducting qubits \cite{Devoret13, Gambetta2017}, and ultracold atoms \cite{Greiner2002, BlochColdAtoms, Bernien2017}, have made many-body quantum simulation a reality.  There have been many successes in the preparation of pure states of interest, such as the Greenberger-Horne-Zeilinger (GHZ) states \cite{PhysRevLett.106.130506}, or in the simulation of quantum dynamics.  {\it Mixed} states, in particular finite temperature states, enable new arenas of interesting physics; perhaps most well-known is the phase diagram of high temperature superconductors, which hosts intriguing phenomena such as ``pseudogaps'' and ``strange metals'' that have eluded a detailed understanding and constitute wide swaths of the finite temperature phase diagram \cite{RevModPhys.78.17}.  

However, the precise preparation of thermal (Gibbs) states poses substantial challenges.  There have been several proposals \cite{PhysRevA.61.022301, qmetropolis, PhysRevLett.103.220502, Kastoryano2016, brandao} which are somewhat formidable for near-term quantum simulators.  For example, quantum Metropolis sampling \cite{qmetropolis} involves subroutines like quantum phase estimation.  On the experimental front, advances in optical lattices of ultracold atoms have enabled finite temperature simulation \cite{esslinger, Gross995, greiner} of specific models such as Bose-Hubbard and Fermi-Hubbard models, but cooling down to low effective temperatures remains challenging.

To address the difficulty of thermal quantum simulation, we borrow from variational schemes for {\it pure} state preparation, which have demonstrated potential in a variety of contexts, from solving classical optimization \cite{farhi_quantum_2014, farhi_quantum_2016, 2017arXiv171205771O, 2018arXiv180810816P, PhysRevA.97.022304} or quantum chemistry problems \cite{vqe1, vqe2, vqe3} to simulating non-trivial phases of matter \cite{variational,2018arXiv180300026H, longrange, zoller18}.  Several variational schemes have been proposed, and the basic idea is to iterate the following steps: (1) prepare on a quantum simulator a wavefunction parameterized by several variables, (2) read out a cost function via single-site measurements, and (3) adjust the variables to decrease the cost function.  This is a hybrid quantum-classical algorithm as the former steps (1,2) involve the quantum simulator while the last step (3) involves classical optimization.  As an example, in the quantum approximate optimization algorithm (QAOA) \cite{farhi_quantum_2014}, the cost function is the energy of a classical Hamiltonian $H_C$, and the variational wavefunction consists of evolving a product state with $H_C$ and a uniform transverse field $H_X$ in an alternating fashion, with each time step as a variational parameter.  Such variational schemes have proven to be simple and efficient, and thus well-suited for near-term quantum simulators.      
\begin{figure}
	\centering
	\includegraphics[width=0.9\linewidth]{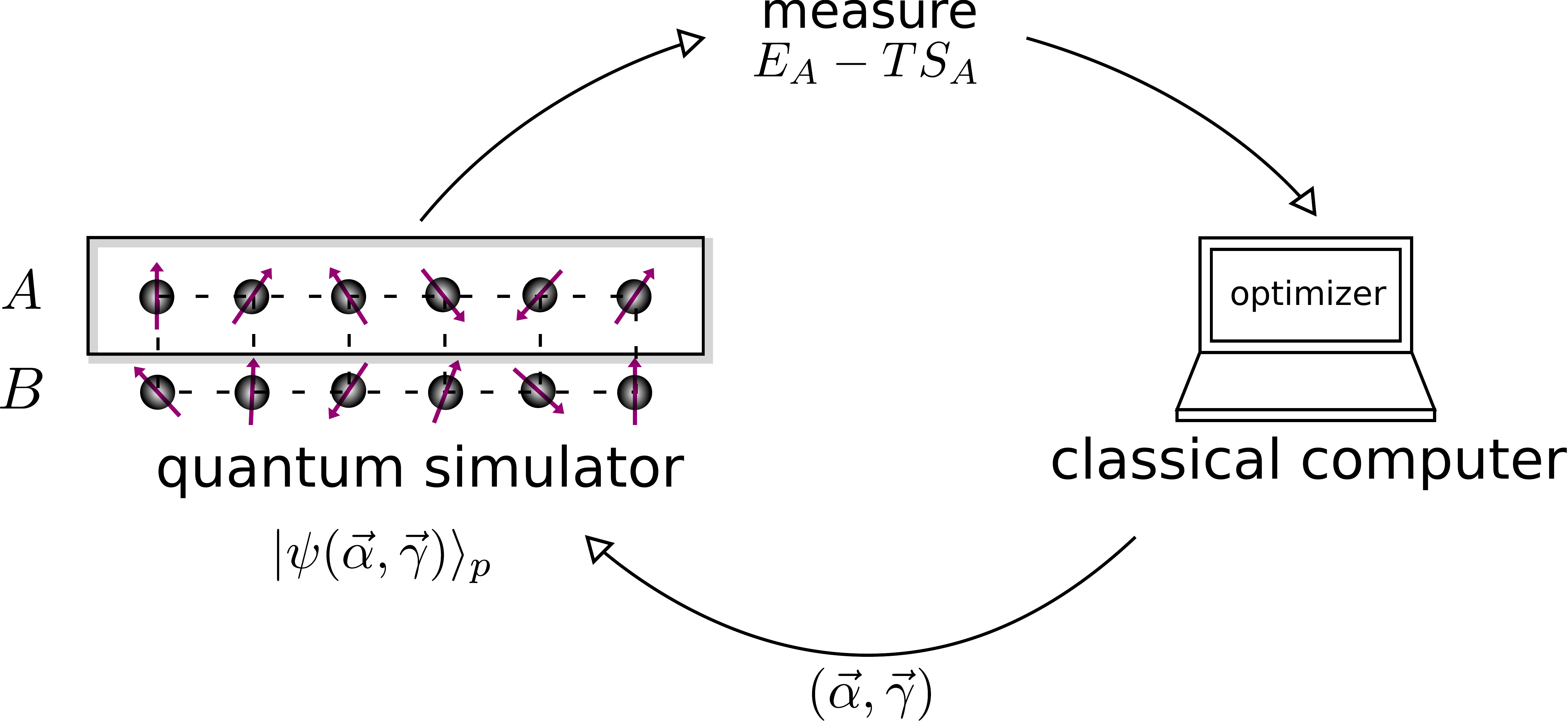}
	\caption{Schematic depiction of the hybrid quantum-classical variational scheme for preparing a thermofield double state at temperature $T$ with respect to a Hamiltonian $H$.  Given a set of parameters $(\vec{\alpha},\vec{\gamma})$, a quantum simulator prepares the wavefunction $| \psi(\vec{\alpha},\vec{\gamma})\rangle_p$ defined in (\ref{eq:target}).  The free energy $E_A - T S_A$, where $E_A$ is the energy of system $A$ and $S_A$ is entanglement entropy, is then measured and updated in a classical computer.  The latter generates a new set of parameters and the process is iterated until the free energy cost function is minimized.}
	\label{fig:cartoon}
\end{figure}

In this work, we propose a variational approach for thermal state preparation that is motivated by QAOA.  Say we wish to prepare the Gibbs state $e^{-\beta H}/Z$ corresponding to inverse temperature $\beta=1/T$ and Hamiltonian $H$ on a system $A$.  Our approach prepares a particular purification of the thermal state called the thermofield double (TFD) state, which is also of interest in the context of the holographic correspondence, where it is dual to a wormhole.  Our variational ansatz for the TFD consists of alternating time evolution between the Hamiltonian of interest $H$ and an ``entangling" Hamiltonian $H_{AB}$ acting jointly on the system and auxiliary degrees of freedom.  In the hybrid quantum-classical setting, the cost function to be minimized for targeting the Gibbs state is the ``free energy'' $F=E_A-TS_A$ of system $A$, where $E_A$ is the energy with respect to $H$ and $S_A$ is the entanglement entropy.       

As proofs of concept, we illustrate the performance of this approach in targeting Gibbs states of classical Ising, free fermion, and spin-1/2 chains in one dimension.  We find that the Gibbs state of the Ising chain at any temperature can be prepared with perfect fidelity using $L/2$ iterations of our protocol, where $L$ is system size.  Furthermore, we show that our protocol for preparing a free fermion thermal state is nearly optimal and consistent with previous theory.  Finally, we show that Gibbs states of quantum spin-1/2 chains can also be prepared efficiently.  Our approach thus allows for the preparation of an interesting class of TFD states and paves an alternative way for quantum simulators to shed light on finite temperature physics.   

{\it Setup:} Consider a Hamiltonian $H_A$ defined on system $A$ and with energies and eigenstates given by $H_A|n\rangle_A = E_n |n\rangle_A$.  Our goal is to prepare the Gibbs state $\rho_\beta = Z^{-1} e^{-\beta H_A}$ by preparing a purification of the Gibbs state called a thermofield double state, defined on an enlarged system.  In particular, let system $B$ consist of identical degrees of freedom as those of $A$.  A TFD is defined on the total system $\mathcal{H}_A\otimes\mathcal{H}_B$ and has the structure
\begin{equation}
|\mathrm{TFD}(\beta)\rangle  = \frac{1}{\sqrt{Z}} \sum_{n} e^{-\beta E_n/2} |n\rangle_A |{\tilde n}\rangle_B,
\end{equation} 

Tracing out system $B$ yields the desired Gibbs state on system $A$.  Note that the TFD is not uniquely defined, as any unitary action on system $B$ will leave $\rho_A$ invariant; we have indicated this ambiguity above with the tilde on the $B$ states.  We define the TFD with the following prescription.  First, we find a Hamiltonian $H_{AB}$ whose ground state is a product state of maximally entangled pairs of $A, B$ degrees of freedom.  We will define this maximally entangled state to be $|\mathrm{TFD}(\beta = 0)\rangle$, since tracing out $B$ yields the maximally mixed (infinite temperature) state on $A$.   Then we define
\begin{equation}
|\mathrm{TFD}(\beta)\rangle \equiv \frac{1}{\sqrt{\cal{N}}}e^{-\beta H_A/2} |\mathrm{TFD}(0)\rangle
\label{eq:target}
\end{equation}
where $\cal{N}$ is a normalization factor.  As one cannot simply evolve in imaginary time on a quantum simulator (see however \cite{2018arXiv180403023M}), it is not obvious how to prepare the finite temperature TFD.

We now define a unitary protocol to prepare $|\mathrm{TFD}(\beta)\rangle$ starting from $|\mathrm{TFD}(0)\rangle$ (which is straightforward to prepare in an experiment given its product state structure).  First we define an operator $H_B$ that is ``dual'' to $H_A$ in the following sense.  The maximal entanglement between $A,B$ of $|\mathrm{TFD}(0)\rangle$ endows it with the property that for every operator $O_A$ supported on $A$, there is a dual operator $O_B$ on $B$ such that $O_A |\mathrm{TFD}(0)\rangle = O_B |\mathrm{TFD}(0)\rangle$ \cite{PhysRevB.94.161112}.  $H_B$ is defined as the dual operator to $H_A$ (in most cases, as will be clear in examples below, it will be of identical form as $H_A$).

Our variational ansatz for the target TFD state is motivated by QAOA and consists of alternating time evolution between $H_A + H_B$ and $H_{AB}$:
\begin{eqnarray}
| \psi(\vec{\alpha},\vec{\gamma})\rangle_p = \prod_{i=1}^{p} e^{i\alpha_i H_{AB}} e^{i\gamma_i (H_A + H_B)/2} |\mathrm{TFD}(0)\rangle 
\label{eq:ansatz}
\end{eqnarray}
Here $\vec{\alpha}, \vec{\gamma}$ are $p$ pairs of variational parameters; the larger the number of variational parameters $p$, the better the ansatz can approximate the target state.  In fact, because $|\mathrm{TFD}(\infty)\rangle$ is the ground state of $H_A + H_B$, and $|\mathrm{TFD}(0)\rangle$ is the ground state of $H_{AB}$, the adiabatic algorithm guarantees that $|\mathrm{TFD}(\infty)\rangle$ can be prepared with perfect fidelity given infinite time.  Trotterizing such an adiabatic protocol yields a unitary circuit of the above form, as $p\rightarrow \infty$.  Intuitively, $|\mathrm{TFD}(\beta =\infty)\rangle$ is the hardest to prepare, and thus we expect the protocol to also work for finite $\beta$.  We will justify this intuition with several examples below.  

The cost function we use for numerical convenience is simply the error in fidelity with respect to the target state:
\begin{equation}
F_p(\vec{\alpha},\vec{\gamma}) = 1-|\langle \mathrm{TFD}(\beta)| \psi(\vec{\alpha},\vec{\gamma})\rangle_p|^2.
\label{eq:cost}
\end{equation}

However, in the hybrid quantum-classical mode of operation, one can use the ``free energy'' cost function defined on system $A$ alone:
\beq
F_A = E_A -TS_A \equiv \mathrm{Tr}[\rho_A H_A] + T \mathrm{Tr} [\rho_A \log \rho_A],
\eeq
where $\rho_A (\vec{\alpha},\vec{\gamma})= \mathrm{Tr}_B | \psi(\vec{\alpha},\vec{\gamma})\rangle \langle \psi(\vec{\alpha},\vec{\gamma})|$.

The relative entropy with respect to the Gibbs state $S(\rho_A || \rho_\beta)$ is non-negative and minimized (zero) when $\rho_A = \rho_\beta$.  Because $S(\rho_A || \rho_\beta)=\beta(F_A -F_{\beta})$, where $F_\beta=-T \log \Tr[e^{-\beta H_A}]$ is the Gibbs free energy, realizing the target Gibbs state is equivalent to minimizing the free energy.  (See appendix for a review of the previous relation and an example using the free energy cost function).  Note that in general, this scheme may yield a TFD state distinct from (\ref{eq:target}), but $\rho_A$ will be the desired Gibbs state. While the energy is straightforward to measure via local observables, approximating the entanglement entropy requires measuring several Renyi entropies, which is non-trivial but achievable in experiments \cite{Islam2015}.  We discuss possible shortcuts to the hybrid feedback scheme in the conclusion.

{\it Ising chain Gibbs state}. --
As a first example, consider a spin $1/2$ chain $A$ of length $L$, with the Ising Hamiltonian
\begin{eqnarray}
H_A = -\sum_{i=1}^{L} Z_{A,i} Z_{A,i+1}  \label{eq:Ising}
\end{eqnarray}
and periodic boundary conditions.  $Z_{A,i}$ denotes the Pauli-Z operator acting on the $i$th site of system $A$.  This system has ferromagnetic order only at strictly zero temperature.  

Consider an auxiliary chain $B$ with identical spin $1/2$ degrees of freedom, and define
\beq
H_{AB} = \sum_{i=1}^{L} X_{A,i}X_{B,i}+Y_{A,i}Y_{B,i} + Z_{A,i}Z_{B,i}
\eeq
In accordance with the protocol, the ground state of $H_{AB}$ is a state $|\mathrm{TFD}(0)\rangle$ that is maximally entangled between $A$ and $B$; it is a product state of Bell pairs (singlets).  The dual Hamiltonian with respect to this state is 
\beq
H_B = -\sum_{i=1}^{L} Z_{B,i} Z_{B,i+1}
\eeq
To approximate the target TFD state $(\ref{eq:target})$ using the ansatz $(\ref{eq:ansatz})$, we numerically find the parameters which minimize the cost function \eqref{eq:cost}.  There are several simplifications that can be made in this example.  Because $Z_{A,i} Z_{B,i}=-1$ for every $i$ is conserved throughout the time evolution, we can make two replacements in the ansatz: $H_{AB}$ can be reduced to $\sum_{i=1}^{L} X_{A,i}X_{B,i}$ and $(H_A + H_B)/2$ can be reduced to $H_A$.    After these replacements, the parameter ranges for $\alpha, \gamma$ can both be chosen to be $[0,\pi/2]$; this is because $\exp(\frac{i\pi}{2} H_A)\propto 1$ and $\exp(\frac{i\pi}{2} \sum_{i=1}^{L} X_{A,i}X_{B,i})\propto \prod X$, which is also a conserved quantity. 

\begin{figure}[tb]
	\centering
	\includegraphics[width=0.9\linewidth]{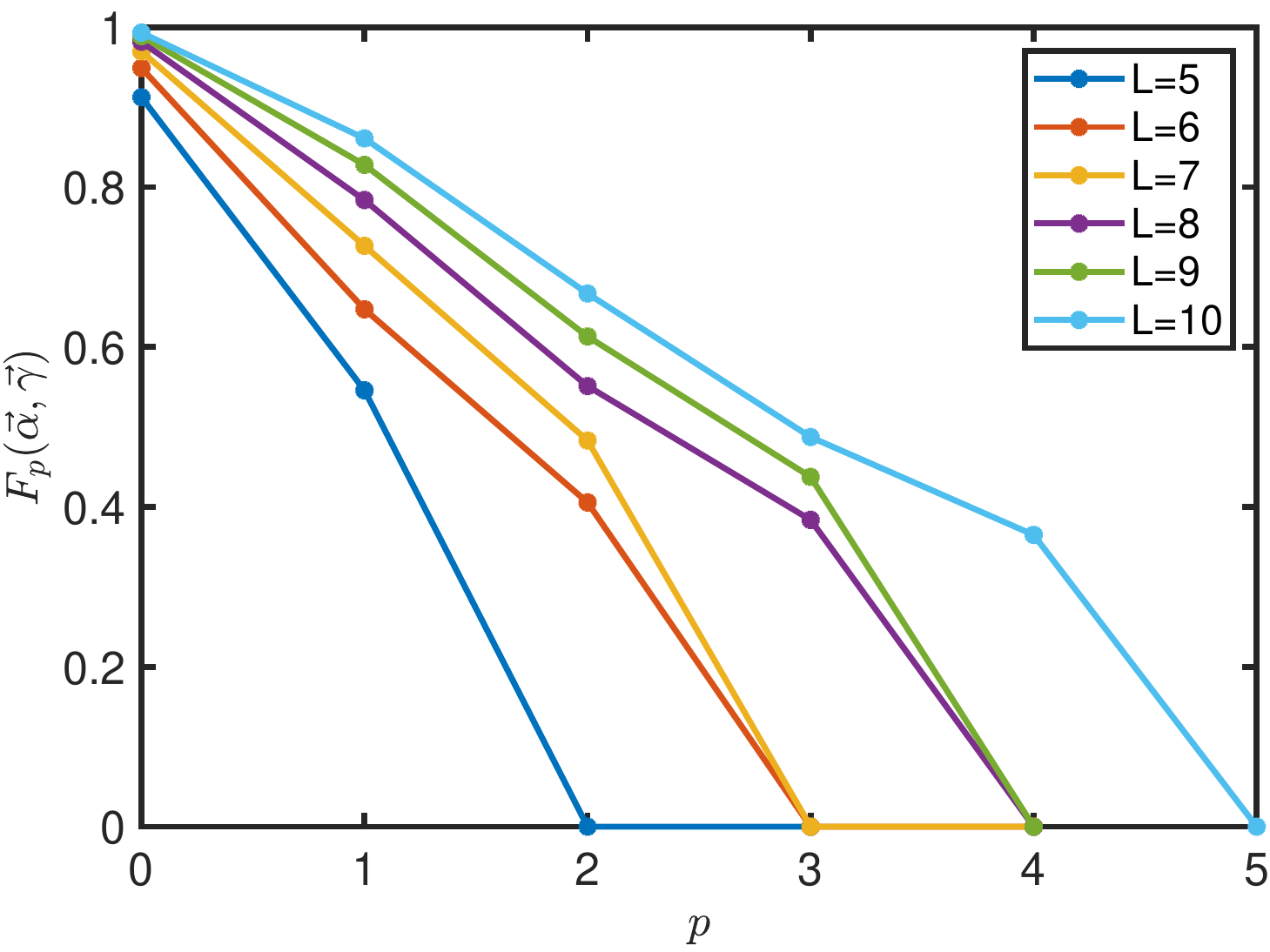}
	\caption{Preparation of Ising chain TFD at $\beta = 2$.  The error in fidelity with respect to the target TFD state is plotted as a function of the number of pairs of variational parameters $p$, for various system sizes.  Perfect fidelity is achieved at $p = \floor{L/2}$, a result which appears to hold for arbitrary temperature.}
	\label{fig:Z+X}
\end{figure}

See Fig. \ref{fig:Z+X} for the results for preparing the Gibbs state at inverse temperature $\beta=2$ on various system sizes $L$ and for different $p$.  It is evident that perfect fidelity can be achieved at $p = L/2$, and we have checked that this holds for other temperatures as well.   This result is closely related to the perfect preparation \cite{2018arXiv180300026H} of ground states of the transverse field Ising model given a QAOA protocol with $p = L/2$ (see appendix for a detailed discussion).  Given a fixed $p$, higher temperature (smaller $\beta$) Gibbs states can be prepared with higher fidelity because they have larger overlap with the initial (infinite temperature) state.  To examine such temperature dependence in more detail, we study a different model. 

{\it Majorana fermion Gibbs state}. --
Let $A$ and $B$ each be a chain of $L$ Majorana modes $\gamma_{A,i}, \gamma_{B,i}$, $i=1...L$.  We consider
\begin{equation}
H_{A} = \sum_{i = 1}^{L}i\gamma_{A,i} \gamma_{A,i+1}, \quad H_{AB} = \sum_{i = 1}^{L}i\gamma_{A,i} \gamma_{B,i}
\end{equation}
with periodic boundary conditions.  $H_A$ has a gapless spectrum consisting of a linearly dispersing Majorana mode, and the ground state of $H_{AB}$ is maximally entangled between $A,B$.   The dual Hamiltonian with respect to this state is  
\begin{equation}
H_{B} = -\sum_{i = 1}^{L}i\gamma_{B,i} \gamma_{B,i+1}
\end{equation}
As in the previous example, $H_{AB}$ satisfies the relation $\exp(\frac{i\pi}{2} H_{AB}) \propto \prod_{i=1}^L (\gamma_{A,i} \gamma_{B,i})$, which is conserved. Therefore $\alpha$ can be chosen to valued in $[0,\pi/2]$.   We do not observe any periodicity in the parameter $\gamma$ and choose a range $[0,2\pi]$ for it. 

\begin{figure}[tb]
	\centering
\includegraphics[width=1.0\linewidth]{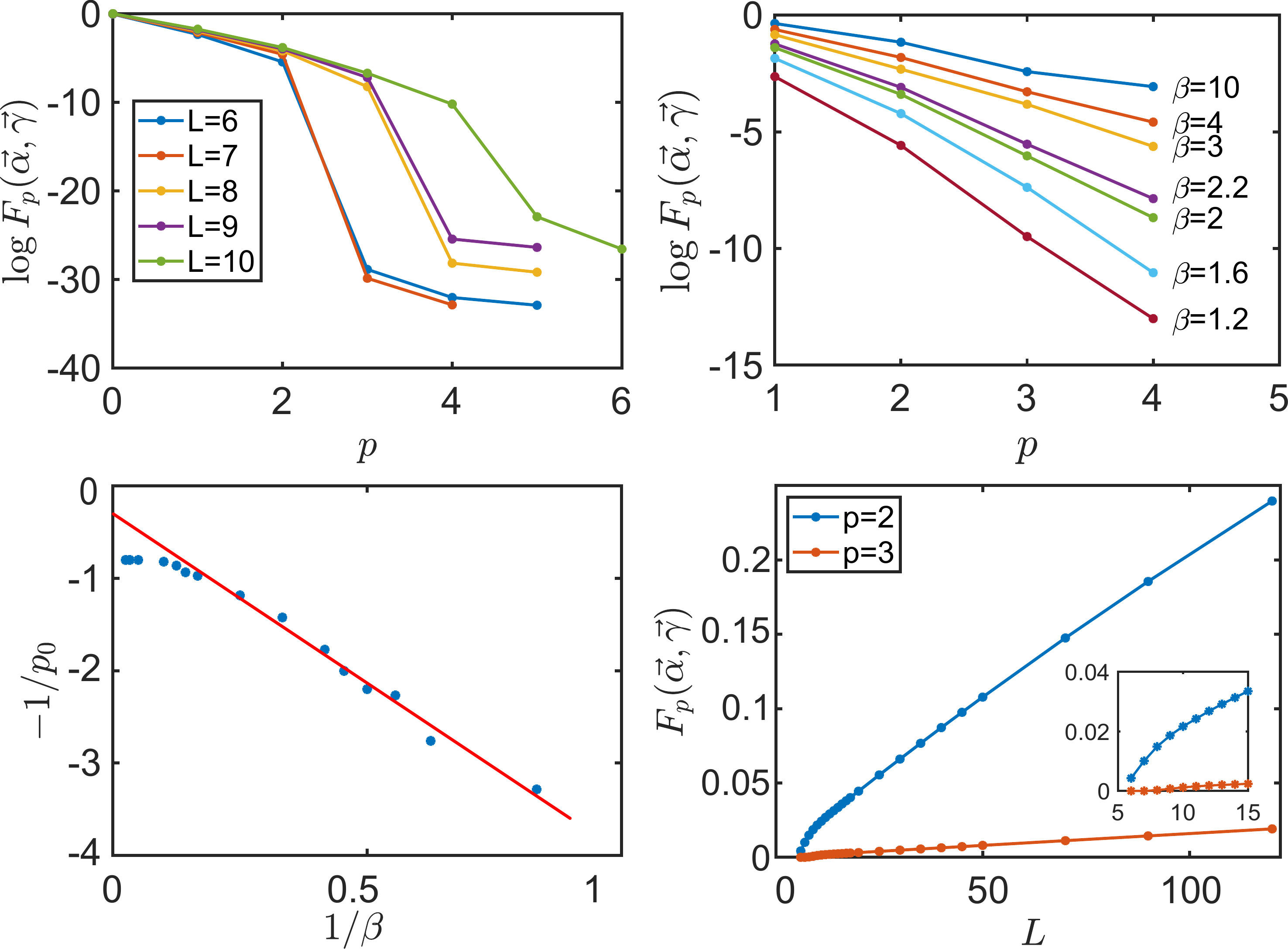}
	\caption{Preparation of Majorana fermion TFD state. (a) Semilog plot of the infidelity for different system sizes at $\beta =2$. (b) Semilog plot of the infidelity of $L=30$ for different temperature values. The infidelity appears to decay exponentially with a slope $-1/p_0$ which depends on temperature.  (c) The dependence of the decay rate $-1/p_0$ on the temperature. The blue dots are slopes from linear fitting of the plots in (b) at each temperature. The red line is the linear fitting of the blue dots for $\beta \in [1,6]$. (d) The dependence of the infidelity at $\beta =2$ on the system size $L$ at $p=2$ and $p=3$. It is evident that the infidelity scales linearly with $L$ for large enough $L$.}
	\label{fig:Maj}
\end{figure}

We show in Fig. \ref{fig:Maj} (a) a semilog plot of the infidelity for targeting the Gibbs state at $\beta = 2$, for various $p,L$ (see Appendix for details of the numerical simulations).   The results suggest that the infidelity decays exponentially with $p$ until around $p_\ast = L/2$ where it drops to nearly zero; $F_p \propto \exp(-p/p_0)$, where the results suggest that $p_0$ is essentially independent of the system size. However, $p_0$ depends strongly on the inverse temperature $\beta$ of the target state. We study this dependence in Fig. \ref{fig:Maj} (b) by varying the temperature while fixing system size to be $L=30$, which is large enough so that $p_\ast \gg 4$ and $1\leq p \leq 4$ is in the regime of exponential decay. The decay rate $-1/p_0$ is extracted by linear fitting for each temperature, and the results for $0.8 \leq \beta \leq 40$ are depicted in Fig. \ref{fig:Maj} (c).  It is evident that $p_0$ is proportional to $\beta$ below a cutoff $\beta_*$, in which the thermal correlation length becomes comparable to system size.  This scaling arises from the fact that the thermal correlation length of the Gibbs state scales as $\xi \propto v\beta$, where $v$ is the velocity of the linearly dispersing Majorana mode.  Assuming that our protocol saturates the light cone growth of correlations (the correlation length grows linearly with circuit depth), we conclude that the depth needed to target $\rho_\beta$ scales linearly with $\beta$.

In Fig.\ref{fig:Maj} (d), we show how the infidelity depends on system size for $p=2,3,4$.  These results in total suggest that for $1<p\ll L/2$, $F_p \propto L \exp(-const*p/\beta)$.  This implies that the depth required to achieve a threshold fidelity with the target state scales as $\beta \log L$.  Such efficiency for creating this free fermion TFD is consistent with \cite{PhysRevB.94.155125}, which showed that all free fermion TFDs at finite temperature are adiabatically connected to the infinite temperature TFD, although the depth of the unitary circuit connecting the two must diverge as $T\rightarrow 0$.  While our $\log L$ dependence prevents us from claiming a strictly finite depth circuit preparation, it is clear from the $p=3$ results in Fig.\ref{fig:Maj} (d) that for all practical purposes the $\log L$ scaling is innocuous.  In this sense, our protocol is nearly optimal.     

{\it Spin chain Gibbs state}. -- Finally, we apply our protocol to target the Gibbs state of the XY spin-1/2 chain $H_A = -\sum_{i=1}^{L} X_{A,i} X_{A,i+1} + Y_{A,i} Y_{A,i+1}$, using 
$H_{AB} = \sum_{i=1}^{L} X_{A,i}X_{B,i}+Y_{A,i}Y_{B,i} + Z_{A,i}Z_{B,i}$ and the dual Hamiltonian $H_B = -\sum_{i=1}^{L} X_{B,i} X_{B,i+1} + Y_{B,i} Y_{B,i+1}$.  We find that our protocol can also achieve high fidelities in this case, although the depth required is larger than that of the Majorana chain.  One possible reason is that the central charge of this critical model is 1 as opposed to the central charge 1/2 of the Majorana example, and more effective degrees of freedom is expected to increase the circuit complexity of the state \cite{Fu2018,2018arXiv180707677G}.    
\begin{figure}[htb]
	\centering
	\includegraphics[width=0.9\linewidth]{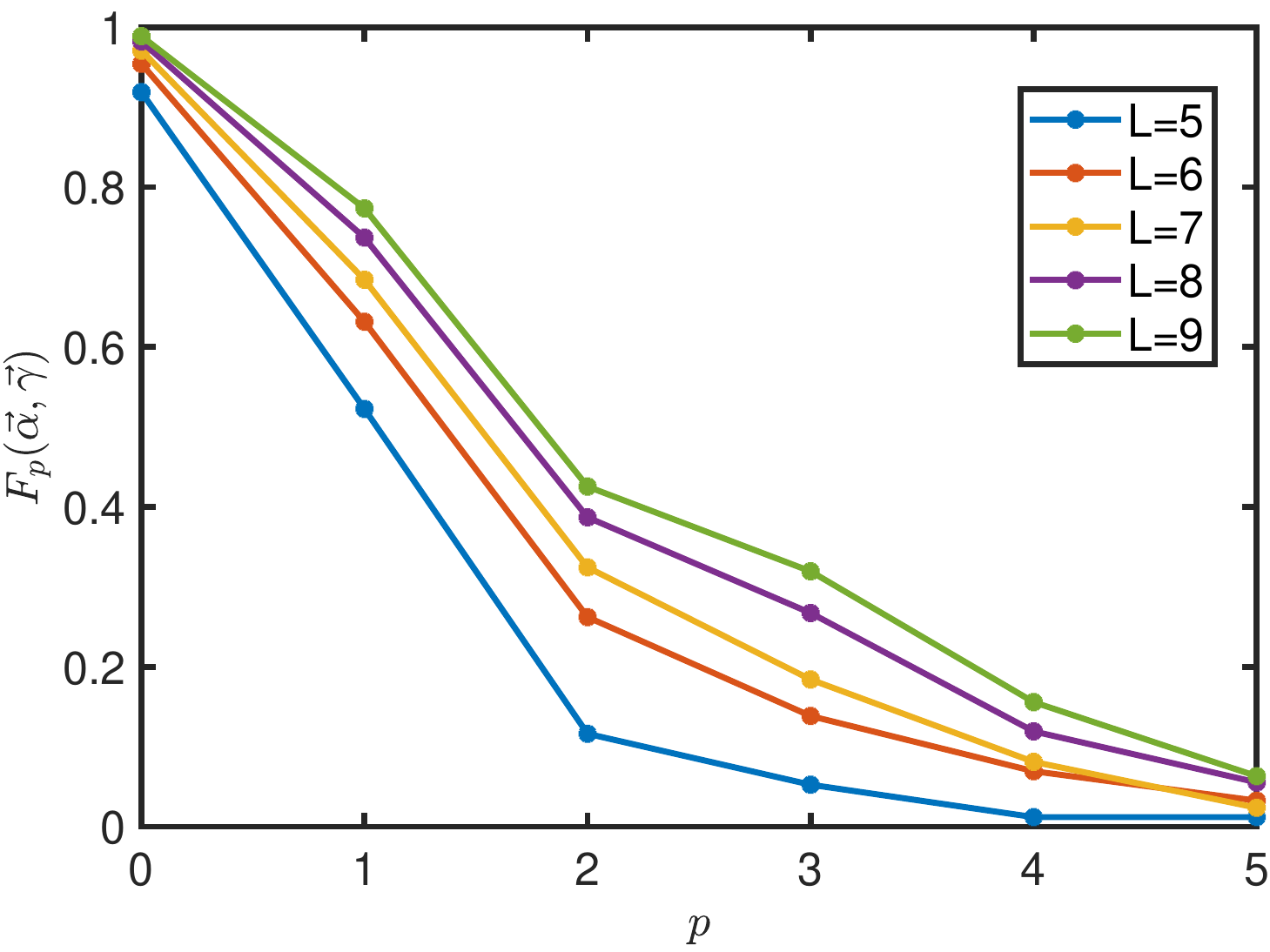}
	\caption{Preparation of the XY spin chain TFD state at $\beta=2$.}
	\label{fig:xyxyz}
\end{figure}

{\it Discussion}. --  We have proposed a variational scheme for preparing thermofield double states that enables the simulation of Gibbs states.  As an example, we have found that Gibbs states of a 1d classical Ising model at any temperature can be prepared perfectly with $L/2$ iterations in our protocol, and Gibbs states of free fermion and (quantum) spin chains can also be prepared efficiently.  Our variational ansatz can be improved further with additional unitary gates; we have presented the simplest scheme in this work.  Likewise, it is interesting to consider shortcuts for measuring the free energy cost function in the hybrid quantum-classical scheme.  For example, instead of optimizing to target the Gibbs state on the entire $A$ system, it may be useful to optimize a local free energy defined on a small subsystem of $A$, which substantially reduces the number of measurements required.  It may then be useful to leverage translation invariance of the protocol to divide the system into small equivalent subsystems, which could serve as the copies needed for Renyi entropy measurements (required to extract the von Neumann entropy of the small subsystem).

In addition to enabling the simulation of finite temperature states, the preparation of TFD states may be useful in itself.  Motivated by the holographic correspondence, in which the TFD is dual to a two-sided black hole, various works \cite{gao1, maldacena2, gao2} have examined how the TFD provides a portal (``traversable wormhole'') for recovering information from one end at a much later time from the other end.  Along these lines, our protocol may also shed light on the circuit complexity required to prepare the TFD state \cite{chapman}, which has also attracted interest due to its holographic significance.  When the Gibbs state undergoes a finite temperature phase transition, the corresponding TFD may no longer be smoothly connected to the infinite temperature TFD \cite{PhysRevB.93.205159, PhysRevB.94.155125}, and it would be interesting to see how this transition is reflected in our protocol.        

Our scheme for variational thermal quantum simulation highlights the fact that finite temperature can be viewed as arising from entanglement with an auxiliary system.  We have presented a concrete setup in which effective temperature of a subsystem can be controlled by simple unitary operations on a joint system.      

{\it Acknowledgments. -- } We thank A. Alhambra for useful discussions. Research at Perimeter Institute
is supported by the Government of Canada through Industry
Canada and by the Province of Ontario through
the Ministry of Research and Innovation. 

\bibliographystyle{JHEP}
\bibliography{reference}
\appendix
\section{Appendix}
\subsection{Relation between Ising TFD and QAOA for transverse field Ising model}
Here we relate the perfect preparation of the Ising chain Gibbs state \eqref{eq:Ising} to the perfect preparation of free fermion states discussed in \cite{2018arXiv180300026H}. For convenience, we restate the relevant Hamiltonians used in our protocol: 
\beq
H_A &=& -\sum_{i=1}^{L} Z_{A,i} Z_{A,i+1},  \\
H_{AB} &=& \sum_{i=1}^{L} X_{A,i}X_{B,i}
\eeq
(Recall that $H_B$ was not necessary for this model.) 
Every $Z_{A,i} Z_{B,i}$ for $i = 1,\dots, L$ is conserved as $-1$, and hence every pair of $A,B$ sites reduces to a two dimensional Hilbert space described by a pseudospin labeled by a tilde.   The problem then reduces to a single chain of pseudospins and the above interactions reduce to 
\beq
 -\sum_{i=1}^{L} \tilde{Z}_{i} \tilde{Z}_{i+1},\qquad  \sum_{i=1}^{L} \tilde{X}_i
\eeq
The initial state $|TFD(0)\rangle$ in this representation is the ground state of the pseudospin paramagnet $\sum_{i=1}^{L} \tilde{X}_i$.  Hence our ansatz exactly reduces to that of the QAOA ansatz for the transverse field Ising model (TFIM) used in \cite{2018arXiv180300026H}.  There, it was conjectured that the above interactions in the QAOA ansatz are sufficient to prepare any state in the phase diagram of the TFIM at depth $p=L/2$.  If we assume a stronger conjecture that any free fermion state with translation symmetry can be prepared at depth $p=L/2$, then it follows that the target TFD state in the problem at hand can be prepared with $p=L/2$.

\subsection{Majorana fermion Gibbs state numerics}
QAOA protocol can be carried out by using the fermionic representation in the same fashion as \cite{PhysRevA.97.022304}. In particular, we apply a Fourier transformation to the Majorana operators
\begin{eqnarray}
\gamma_{2j-1} = \frac{1}{\sqrt{L}} \sum_{k=0}^{L-1} c_k e^{i\frac{2\pi}{L}kj} +c^\dagger_k e^{-i\frac{2\pi}{L}kj}\\
\gamma_{2j} = \frac{-i}{\sqrt{L}} \sum_{k=0}^{L-1} c_k e^{i\frac{2\pi}{L}kj} -c^\dagger_k e^{-i\frac{2\pi}{L}kj}
\end{eqnarray}
The Hamiltonians of $A$ chain and B chain and interchain take the form
\begin{align}
H_{AB} = & \sum_{k=1}^{L/2-1} (2c_k^\dagger c_k + 2c_{-k}^\dagger c_{-k} -2)\nonumber\\
& + 2c_0^\dagger c_0 -1 +2 c^\dagger_{L/2} c_{L/2} -1 \\
H_s = & \sum_{k=1}^{L/2-1} -2 \sin (\frac{2\pi}{L} k) (c^\dagger_{k} c^\dagger_{-k} -c_k c_{-k})
\end{align}
where $H_s = (H_A+H_B)/2$. In QAOA protocol, the initial state is the ground state of $H_{AB}$, whose eigenvalue under each number operator $c^\dagger_k c_k$ is $0$. Note that both $H_{AB}$ and $H_s$ preserve the fermionic parity. Therefore in each subspace created by $c_k^\dagger$ and $c_{-k}^\dagger$, we only need to consider empty state and double occupied state, effectively reduce the Hilbert space to the tensor product of $L/2$ decoupled two dimensional systems, namely the problem can be recast as
\begin{align}
\tilde{H}_{AB} &= \sum_{k=0}^{L/2-1} 2 Z_k\\
\tilde{H_s} & = \sum_{k=0}^{L/2-1} 2 \sin(\frac{2\pi}{L}k) Y_k
\end{align} 
We will drop the tilde from now on. To proceed with QAOA, we set the angle range to be $\alpha_i \in [0,\pi/2]$ and $\gamma_i \in [0,2\pi]$ due to the same reason that $\exp(i\pi/2 H_{AB}) $ is the fermion parity operator, which is conserved and that no obvious periodicity can be found in $\gamma_i$.

\section{Optimization using free energy cost function}

The relative entropy between the reduced density matrix $\rho_A$ on the $A$ subsystem and the target Gibbs state $\rho_\beta$ is
\beq
S(\rho_A || \rho_\beta)&=&\Tr \rho_A \log \rho_A -\Tr \rho_A \log \rho_\beta \\
&=&-S_A + \beta E_A+\log Tr e^{-\beta H_A} \\
&=&\beta(F_A - F_\beta)
\eeq 
where $F_A=E_A -TS_A $ and $F_\beta=-T \log \Tr[e^{-\beta H_A}]$ are free energies.

\begin{figure}[htb]
	\centering
	\includegraphics[width=0.9\linewidth]{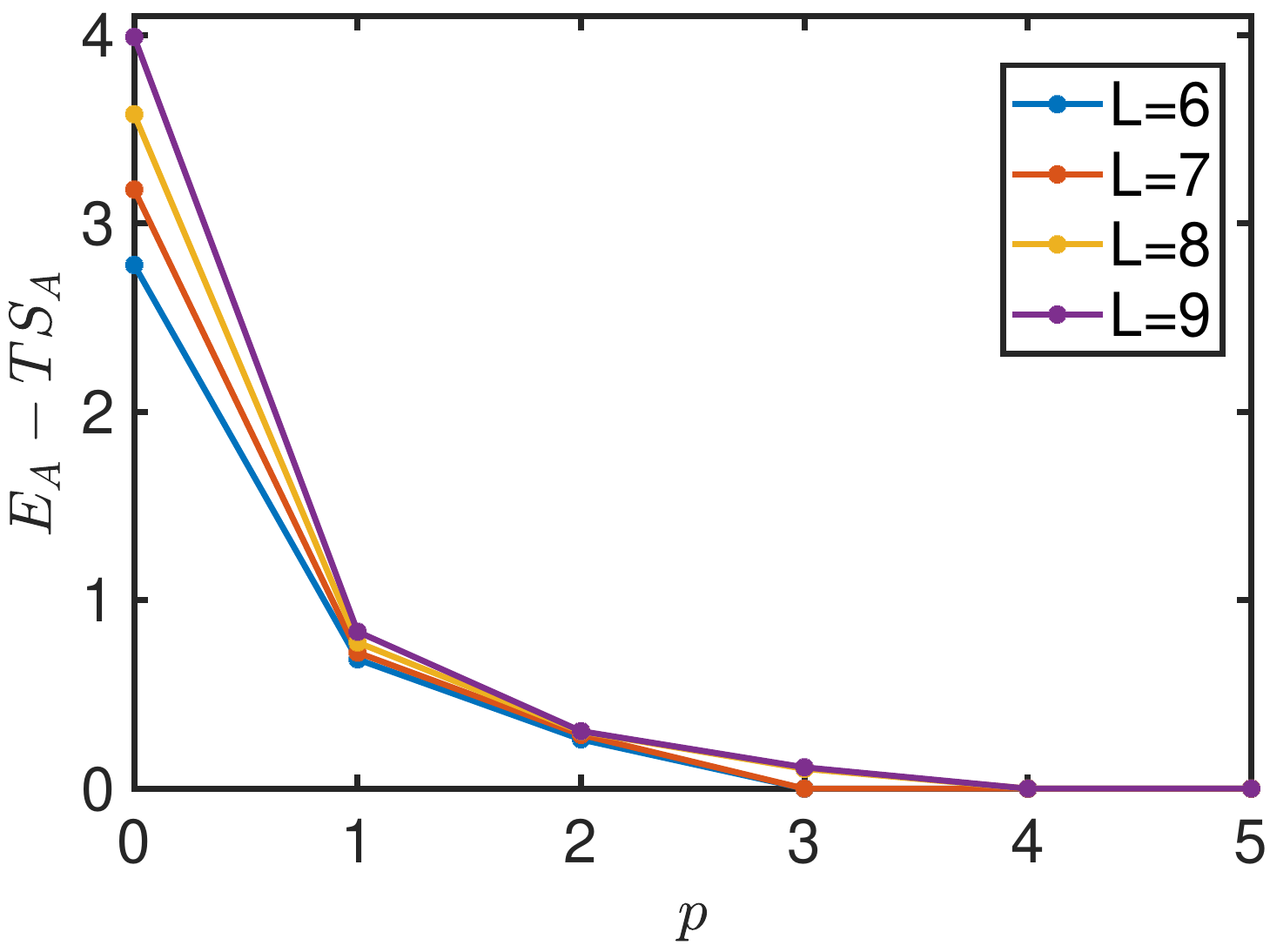}
	\caption{Preparation of Ising Gibbs state at $\beta=1$ using the free energy cost function. We have subtracted the constant $\beta F_\beta$}
	\label{fig:ets}
\end{figure}

We can minimize this cost function and demonstrate this in the example of Ising chain Gibbs state, shown in Fig.\ref{fig:ets}. Note that we subtract the minimal value $\beta F_\beta$ in the plot. Evidently, we again reach the target Gibbs state after $p = L/2$ steps. Note however that we don't necessarily reach the particular TFD state defined in \eqref{eq:target}; it only matters that for the subsystem $\rho_A = \rho_\beta$.

\end{document}